\documentclass[nofootinbib,prd,showpacs]{revtex4}
\usepackage{amsmath}
\usepackage{subfigure}
\usepackage{amsfonts}
\usepackage{hyperref}
\usepackage{amssymb}
\usepackage{graphicx}
\graphicspath{ {figs/} }

\usepackage[latin1]{inputenc}
\usepackage[english]{babel}
\usepackage{color,xcolor}

\definecolor{purple}{rgb}{1,0,1}
\definecolor{lime}{HTML}{A6CE39} 

\definecolor{lime}{HTML}{A6CE39}

\begin{document}
\title{
Source of black bounces in general relativity}
\author{Manuel E. Rodrigues}
\email{esialg@gmail.com}
\affiliation{Faculdade de Ci\^{e}ncias Exatas e Tecnologia, 
Universidade Federal do Par\'{a}\\
Campus Universit\'{a}rio de Abaetetuba, 68440-000, Abaetetuba, Par\'{a}, 
Brazil}
\affiliation{Faculdade de F\'{\i}sica, Programa de P\'{o}s-Gradua\c{c}\~ao em 
F\'isica, Universidade Federal do 
 Par\'{a}, 66075-110, Bel\'{e}m, Par\'{a}, Brazil}
\author{Marcos V. de S. Silva}
\email{marco2s303@gmail.com}
\affiliation{Faculdade de F\'{\i}sica, Programa de P\'{o}s-Gradua\c{c}\~ao em 
F\'isica, Universidade Federal do 
 Par\'{a}, 66075-110, Bel\'{e}m, Par\'{a}, Brazil}
\date{today; \LaTeX-ed \today}
\begin{abstract}
Black bounces are spacetimes that can describe, depending on certain parameters, black holes or wormholes. In this work, we use a method to obtain the matter content that generates black bounce solutions in general relativity. The method is constructed in a general way, and as models, we apply it to the Simpson--Visser black bounce solution and the Bardeen-type black bounce solution. We obtain that these metrics are solutions of Einstein's equations when we consider the coupling of the gravitational interaction with a phantom scalar field with a nonlinear electrodynamics. The presence of the phantom scalar field is linked to the fact that this type of solution violates the null energy condition. We analyze separately the energy conditions associated with the stress-energy tensor for the scalar field and for the electromagnetic field.
\end{abstract}
\maketitle
\def\HMS{{\scriptscriptstyle{HMS}}}
\section{Introduction}
\label{S:intro}

From the classical point of view, general relativity is the theory that, in the simplest way, best describes gravitational interaction \cite{din,wal}. This theory was able to solve existing problems and predict new phenomena, such as the bending of light and the existence of gravitational waves \cite{Einstein:1915bz,Kraniotis:2003ig,Will:2018mcj,Crispino:2019yew,LIGOScientific:2016aoc,LIGOScientific:2018mvr,LIGOScientific:2020ibl,LIGOScientific:2021psn}. Mathematically, general relativity is described through the Einstein equations, a set of nonlinear, second-order differential equations coupled in the components of the metric tensor \cite{wal,Chandrasekhar:1985kt}. Depending on the imposed conditions, such as symmetry or matter content, Einstein equations have different solutions. One of the best known solutions is the Schwarzschild metric. The Schwarzschild line element can be used to study spacetime around static stars. This solution also describes the simplest model of a black hole that exists, since it has no spin or charge \cite{Chandrasekhar:1985kt}. From an astrophysical point of view, black holes are compact objects with a strong gravitational field, which not even light can escape \cite{wal}. This strong field allows us to test general relativity more precisely, allowing to measure small deviations. Currently, these experiments are the detection of gravitational waves and images of black holes \cite{LIGOScientific:2016aoc,EventHorizonTelescope:2019dse}.

Another interesting solution to Einstein equations is the wormhole. This object is distinguished by the possibility of connecting two different points of the same universe, or from different universes, through a tunnel, the throat \cite{book}. The first wormhole solution was proposed by Einstein and Rose and is basically the maximum extension of the Schwarzschild solution \cite{Einstein:1935tc}. Despite connecting two distinct regions, the Einstein--Rosen solution is not traversable, which means that, no particle can cross through the throat. The first traversable wormhole solution was proposed by Ellis and Bronnikov, and later studied again by Morris and Thorne \cite{Ellis:1973yv,Bronnikov:1973fh,Morris:1988cz}. Interestingly, wormholes can mimic the ringdown of black holes \cite{Cardoso:2016rao}. Thus, the signal of a gravitational wave is not a definitive proof of the presence of an event horizon. One problem that arises with these solutions is that exotic matter content is required to maintain them. This exotic matter violates energy conditions, such as phantom scalar fields and nonlinear electrodynamics \cite{Lobo:2005us,Lobo:2007zb,Visser:2003yf,Barcelo:2000zf,Bronnikov:2018vbs,Bronnikov:2017sgg,Alcubierre:2017pqm}. However, solutions have recently emerged with a more viable material content, considering for example a Dirac field \cite{Blazquez-Salcedo:2020czn,Bolokhov:2021fil,Konoplya:2021hsm,Churilova:2021tgn}. In the literature there is a very extensive number of works involving wormholes, from geodesics to the stability of these solutions \cite{Bronnikov:2012ch,Bronnikov:2013coa,Bronnikov:2001ils,Lobo:2005yv,Visser:1989kh,Lemos:2003jb,Jusufi:2017mav,Cramer:1994qj}.

Nonlinear electrodynamics is also involved with another type of solution, known as regular black hole. These solutions are characterized by the fact that they have event horizons but do not have singularities like traditional black holes, such as Schwarzschild \cite{Ansoldi:2008jw}. The first regular solution was proposed by James Bardeen \cite{Bardeen}. However, were Beato and Garcia who showed that this solution arises from Einstein equations coupled with nonlinear electrodynamics \cite{Beato1}. These solutions have interesting properties, such as the fact that photons do not follow geodesics or changes in the thermodynamics of these solutions \cite{NED4,Rodrigues:2022qdp}. Using nonlinear electrodynamics, it is also possible to construct solutions with multiple horizons \cite{Rodrigues:2020pem}. As in the case of wormholes, the study of regular solutions has been extensively explored in recent decades \cite{Bronnikov:2017tnz,Zaslavskii,Rodrigues:2015,Rodrigues:2017,Rodrigues:2018,Rodrigues:2019,Silva:2018,Junior:2020,NED5,neves,toshmatov,Bernar:2019xxf}.

Recently, Simpson and Visser proposed a new type of regular solution that, depending on the choice of parameter, can become the Schwarzschild solution, a regular black hole, an one-way traversable wormhole, and a two-way traversable wormhole \cite{Simpson:2018tsi}. This type of solution is known as a black bounce. This solution has a throat at $r=0$ and the area of the event horizon has no dependence on the parameter of the solution. In addition to the Simpson-Visser solution, there are other black bounce models \cite{Lobo:2020ffi,Junior:2022zxo,Rodrigues:2022mdm,Huang:2019arj}. However, like the previous solution, most of these metrics were not proposed as solutions to Einstein equations. Even without knowing the material content of these solutions, several properties can be analyzed \cite{Yang:2022ryf,Rodrigues:2022rfj,LimaJunior:2022zvu,Ghosh:2022mka,Zhang:2022nnj,Yang:2022xxh,Tsukamoto:2022vkt,Bambhaniya:2021ugr,Xu:2021lff,Yang:2021cvh,Guerrero:2021ues,Tsukamoto:2021caq,Franzin:2021vnj,Islam:2021ful,Cheng:2021hoc,Zhou:2020zys,Tsukamoto:2020bjm,Nascimento:2020ime,Lobo:2020kxn}. However, there are other properties that can only be studied when the source of matter is known. Pedro and Bronnikov, in different works, proposed ways to obtain this material content considering the nonlinear electrodynamics with a phantom scalar field \cite{Canate:2022gpy,Bronnikov:2021uta,Bronnikov:2022bud}.

Phantom scalar fields are commonly associated with wormholes \cite{Lobo:2005us}. This type of source is capable of generating wormhole solutions even with minimal coupling \cite{Ellis:1973yv}. These fields are distinguished by the presence of a negative energy density, thus violating the known energy conditions. In general, solutions of this type end up having a series of complications such as instability or even poorly defined thermodynamics \cite{Bronnikov:2012ch}. Another type of solution, which are similar to the black bounces, that can be generated by the phantom scalar field is black universe \cite{Bronnikov:2006fu,Bronnikov:2011zz,Bronnikov:2015kea,Chataignier:2022yic,Bolokhov:2015qaa,Bronnikov:2016xvj}. In addition to these facts, the phantom scalar field can still be treated as a candidate for dark energy, further increasing its relevance. Still, on the phantom scalar field, we can have black hole solutions, also known as phantom black holes. This was first proposed by Bergmann and Leipnik, in 1957 \cite{Bergmann:1957zza}.  Later, other authors consider this type of source to find more solutions \cite{Gao:2006iw,Azreg-Ainou:2011gcq,Jardim:2012se}. There are also regular versions of these solutions \cite{Bronnikov:2005gm,Eiroa:2013nra,Huang:2016qnl}. Some phantom black holes, known as cold black holes, have zero temperature \cite{Bronnikov:1997js,Bronnikov:2006qj,Bronnikov:2006fa}.

The structure of this paper is organized as follows. In Sec. \ref{S:Fieldeq}, we give the motivation why we should use the phantom scalar field with nonlinear electrodynamics to generate black bounce solutions. We also use Einstein equations and build, in a general form, the formalism that will be used to obtain the material content of the solutions.
In the Sec. \ref{S:SV} and Sec. \ref{S:BT}, using the method constructed in the section before, we obtain the material content of the Simpson--Visser solution and the Bardeen-type black bounce. Section \ref{S:EC} is dedicated to the study the energy conditions for each field that makes up the source of the solutions. Our conclusions and perspectives are present in Sec. \ref{S:conclusion}.

 We adopt the metric signature $(+,-,-,-)$.
 We shall work in geometrodynamics units where $G=\hbar=c=1$. 

\section{General Solution}
\label{S:Fieldeq}
Black bounces are structures that have a throat covered by an event horizon. These compact objects interpolate between a regular black hole and a wormhole.

The Simpson--Visser solution describes a black bounce, and is given by the line element \cite{Simpson:2018tsi}
\begin{equation}
    ds^2=f(r)dt^2-f(r)^{-1}dr^2-\Sigma(r)^2\left(d\theta^2+\sin^2\theta d\varphi^2\right),\label{lineel}
\end{equation}
with
\begin{equation}
    f(r)=1-\frac{2m}{\sqrt{r^2+a^2}}, \quad \mbox{and} \quad \Sigma(r)=\sqrt{r^2+a^2}.\label{SV-model}
\end{equation}

If we consider general relativity, the black bounce model proposed by Simpson--Visser does not satisfy Einstein equations in a vacuum. In fact, we can interpret the matter content as being an anisotropic fluid given by
\begin{eqnarray}
    \rho &=& -\frac{a^2\left(\sqrt{r^2+a^2}-4m\right)}{8\pi\left(r^2+a^2\right)^{5/2}},\label{densout}\\
    p_1&=& -\frac{a^2}{8\pi\left(r^2+a^2\right)^{2}},\label{prout}\\
    p_2 &=& -\frac{a^2\left(\sqrt{r^2+a^2}-m\right)}{8\pi\left(r^2+a^2\right)^{5/2}}\label{ptout},
\end{eqnarray}
where $t$ is the timelike coordinate, and
\begin{eqnarray}
    \rho &=& -\frac{a^2}{8\pi\left(r^2+a^2\right)^{2}},\label{densin}\\
    p_1&=& -\frac{a^2\left(\sqrt{r^2+a^2}-4m\right)}{8\pi\left(r^2+a^2\right)^{5/2}},\label{prin}\\
    p_2 &=& -\frac{a^2\left(\sqrt{r^2+a^2}-m\right)}{8\pi\left(r^2+a^2\right)^{5/2}}\label{ptin},
\end{eqnarray}
where $t$ is the spacelike coordinate. The quantities $\rho$, $p_1$, and $p_2$ are the components of the stress-energy tensor
\begin{equation}
    {T^{\mu}}_{\nu}=\mbox{diag} \left[\rho,- p_1, -p_2, -p_2\right],
\end{equation}
where $t$ is the timelike coordinate, and
\begin{equation}
    {T^{\mu}}_{\nu}=\mbox{diag} \left[-p_1,\rho, -p_2, -p_2\right],
\end{equation}
where $t$ is the spacelike coordinate.

A first attempt to build a source for this solution would be to consider nonlinear electrodynamics. However, the stress-energy tensor for nonlinear electrodynamics has the symmetry ${T^{0}}_0={T^{1}}_1$ \cite{NED4}, and we see from equations \eqref{densout}-\eqref{prout} that ${T^{0}}_0\neq {T^{1}}_1$. This implies that the Simpson--Visser metric cannot be interpreted as a solution of Einstein equations in the presence of nonlinear electrodynamics. This type of behavior is not unique to the Simpson--Visser solution.

To solve that problem, let us consider the theory described by the action
\begin{equation}
    S=\int d^4x\sqrt{-g}\left[R-2\kappa^2\left(\epsilon g^{\mu\nu}\partial_\mu\phi\partial_\nu\phi-V\left(\phi\right)\right)+2\kappa^2L(F)\right],\label{action}
\end{equation}
where $R$ is the Ricci scalar, $g_{\mu\nu}$ are the components of the metric tensor, $g$ is the determinant of the metric, $\phi$ is a scalar field, $V\left(\phi\right)$ is the potential related with the scalar field and the electromagnetic Lagrangian $L(F)$ is an arbitrary function of the electromagnetic scalar $F=F^{\mu\nu}F_{\mu\nu}/4$. Here, $\epsilon=\pm 1$. To $\epsilon=+ 1$ we have a usual scalar field, and to $\epsilon=-1$ we have a phantom scalar field \cite{Bronnikov:2006qj}. 
 
The field equations related with the action \eqref{action} are
\begin{eqnarray}
 &&   \nabla_\mu \left[L_F F^{\mu\nu}\right]=\frac{1}{\sqrt{-g}}\partial_\mu \left[\sqrt{-g}L_F F^{\mu\nu}\right]=0,\\
 &&   2\epsilon \nabla_\mu \nabla^\mu\phi=-\frac{dV(\phi)}{d\phi},\\
   && R_{\mu\nu}-\frac{1}{2}g_{\mu\nu}R=\kappa^2T^{\phi}_{\mu\nu}+\kappa^2T^{EM}_{\mu\nu},
\end{eqnarray}
where $L_F=\partial L/\partial F$, $R_{\mu\nu}$ is the Ricci tensor, $T^{\phi}_{\mu\nu}$ is the stress-energy tensor of the scalar field, and $T^{EM}_{\mu\nu}$ is the stress-energy tensor of the electromagnetic field. The stress-energy tensors $T^{\phi}_{\mu\nu}$ and $T^{EM}_{\mu\nu}$ are given by
\begin{eqnarray}
    T^{\phi}_{\mu\nu}&=&2\epsilon \partial_\mu\phi\partial_\nu\phi-g_{\mu\nu}\left(\epsilon g^{\alpha\beta}\partial_\alpha \phi \partial_\beta \phi-V(\phi)\right),\\
    T^{EM}_{\mu\nu}&=&g_{\mu\nu}L(F)-L_F {F_\mu}^\alpha F_{\nu \alpha}.
\end{eqnarray}

The line element that describes a general black bounce spacetime is written as \cite{Lobo:2020ffi}
\begin{equation}
    ds^2=f(r)dt^2-f(r)^{-1}dr^2-\Sigma^2 (r) \left(d\theta^2+\sin^2\theta d\varphi^2\right).
\end{equation}

We will consider only magnetic charged solutions. So that, the only nonzero component of the Maxwell-Faraday tensor, $F_{\mu\nu}$, is
\begin{equation}
    F_{23}=q \sin \theta,\label{f23}
\end{equation}
and the electromagnetic scalar is
\begin{equation}
    F(r)=\frac{q^2}{2\Sigma^4}.\label{EMScalar}
\end{equation}

The equations of motion to the gravitational field and the scalar field are
\begin{eqnarray}
    -\frac{f'(r) \Sigma '(r)}{\Sigma (r)}-\frac{f(r) \Sigma '(r)^2}{\Sigma
   (r)^2}-\frac{2 f(r) \Sigma ''(r)}{\Sigma (r)}+\frac{1}{\Sigma (r)^2}=\kappa ^2 L(r)+\kappa ^2 \epsilon
   f(r) \phi '(r)^2+\kappa ^2 V(r),\label{eqGR1}\\
   -\frac{f'(r) \Sigma '(r)}{\Sigma (r)}-\frac{f(r) \Sigma '(r)^2}{\Sigma
   (r)^2}+\frac{1}{\Sigma (r)^2}=\kappa ^2
   L(r)-\kappa ^2 \epsilon  f(r) \phi '(r)^2+\kappa ^2 V(r),\label{eqGR2}\\
   -\frac{f'(r) \Sigma '(r)}{\Sigma (r)}-\frac{f''(r)}{2}-\frac{f(r) \Sigma
   ''(r)}{\Sigma (r)}=\kappa ^2
   L(r)-\frac{\kappa ^2 q^2 L_F(r)}{\Sigma (r)^4}+\kappa ^2 \epsilon  f(r) \phi '(r)^2+\kappa ^2 V(r),\label{eqGR3}\\
   -2 \epsilon  \left(f'(r) \phi '(r)+f(r) \phi ''(r)\right)-\frac{4
   \epsilon  f(r) \Sigma '(r) \phi '(r)}{\Sigma (r)}=-\frac{V'(r)}{\phi
   '(r)}.\label{eqphi}
\end{eqnarray}

From the field equations, we find that
\begin{equation}
    \phi'(r)^2=-\frac{\Sigma ''(r)}{\kappa ^2 \epsilon  \Sigma (r)},\quad \mbox{or} \quad \phi'(r)=\frac{i }{\kappa  \sqrt{\epsilon } }\sqrt{\frac{\Sigma ''(r)}{\Sigma (r)}}.\label{scalarderi}
\end{equation}
Usually, $\Sigma ''(r)/\Sigma (r)>0$, so that, to guarantee that the scalar field is real, we need $\epsilon=-1$. The scalar field is real only to a phantom scalar field. If we calculate the null energy condition, in regions where $f(r)>0$, one of the inequalities is
\begin{equation}
    NEC_1 \Longleftrightarrow -\frac{2f\Sigma ''}{\kappa^2 \Sigma}\geq 0.
\end{equation}
As $\Sigma ''(r)/\Sigma (r)>0$, the null energy condition is violated. However, if $\Sigma ''(r)/\Sigma (r)<0$, the null energy condition is satisfied and we have a real scalar field to $\epsilon=1$. Based on these points, we can elaborate the following theorem:

\textit{\textbf{Theorem}: For any black bounce solution that arises from coupling the gravitational field with a scalar field and nonlinear electrodynamics, the scalar field must necessarily be phantom if $\Sigma ''(r)/\Sigma (r)>0$, i.e., if the inequality $NEC_1$ is violated. In this case, $\epsilon=-1$.}

From equations \eqref{eqGR1}-\eqref{eqphi} we find
\begin{eqnarray}
   L(r)&=& -\frac{f'(r) \Sigma '(r)}{\kappa ^2 \Sigma (r)}-\frac{f(r) \Sigma
   '(r)^2}{\kappa ^2 \Sigma (r)^2}+\epsilon  f(r) \phi
   '(r)^2+\frac{1}{\kappa ^2 \Sigma (r)^2}-V(r),\label{Lgeneral}\\
   L_F(r)&=&\frac{\Sigma (r)^4 f''(r)}{2 \kappa ^2 q^2}-\frac{f(r) \Sigma (r)^2
   \Sigma '(r)^2}{\kappa ^2 q^2}+\frac{f(r) \Sigma (r)^3 \Sigma
   ''(r)}{\kappa ^2 q^2}+\frac{2 \epsilon  f(r) \Sigma (r)^4 \phi
   '(r)^2}{q^2}+\frac{\Sigma (r)^2}{\kappa ^2 q^2},\label{LFgeneral}\\
   V'(r)&=&\frac{2 \phi '(r) \left(\epsilon  \Sigma (r) f'(r) \phi '(r)+2 \epsilon
    f(r) \Sigma '(r) \phi '(r)+\epsilon  f(r) \Sigma (r) \phi
   ''(r)\right)}{\Sigma (r)}.\label{Vgeneral}
\end{eqnarray}

To solve \eqref{scalarderi} and to obtain \eqref{Lgeneral}-\eqref{Vgeneral}, we need to specify $f(r)$ and $\Sigma(r)$.


\section{Simpson--Visser Solution}\label{S:SV}
The Simpson--Visser model is given by the line element \eqref{lineel} with \eqref{SV-model}. Here, we will consider that the parameter $a=q$ is the magnetic charge.

\begin{figure}
    \centering
    \includegraphics[scale=0.7]{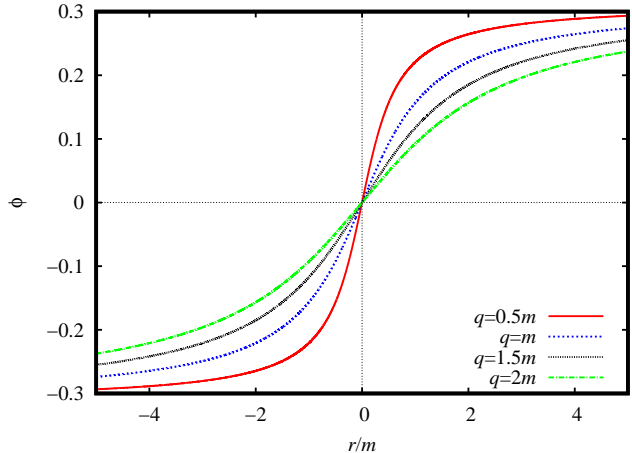}
    \caption{Behavior of the scalar field as a function of the radial coordinate to different values of charge.}
    \label{fig:Phi}
\end{figure}

We find that the scalar field, the potential, and the electromagnetic quantities are
\begin{eqnarray}
    \phi(r)&=&\frac{\tan ^{-1}\left(\frac{r}{q}\right)}{\kappa},\label{phi-SV}\\
    V(r)&=&\frac{4 m q^2}{5 \kappa ^2 \left(q^2+r^2\right)^{5/2}},\label{V-SV}\\
    L(r)&=&\frac{6 m q^2}{5 \kappa ^2 \left(q^2+r^2\right)^{5/2}},\label{L-SV}\\
    L_F(r)&=&\frac{3 m}{\kappa ^2 \sqrt{q^2+r^2}}.\label{LF-SV}
\end{eqnarray}
In Fig. \ref{fig:Phi} we show the behavior of the scalar field in terms of the radial coordinate for different charge values. The field is always positive for positive radial coordinate values and always negative for negative radial coordinate values. We thus see that the scalar field is not symmetric $r\rightarrow-r$.

The electromagnetic quantities must obey the relation
\begin{equation}
    L_F-\frac{\partial L}{\partial F}= L_F-\frac{\partial L}{\partial r}\left(\frac{\partial F}{\partial r}\right)^{-1}=0.\label{EMbound}
\end{equation}
If we substitute the expressions \eqref{L-SV} and \eqref{LF-SV} in equation \eqref{EMbound}, we verify that it is identically satisfied.

\begin{figure}
    \centering
    \includegraphics[scale=0.7]{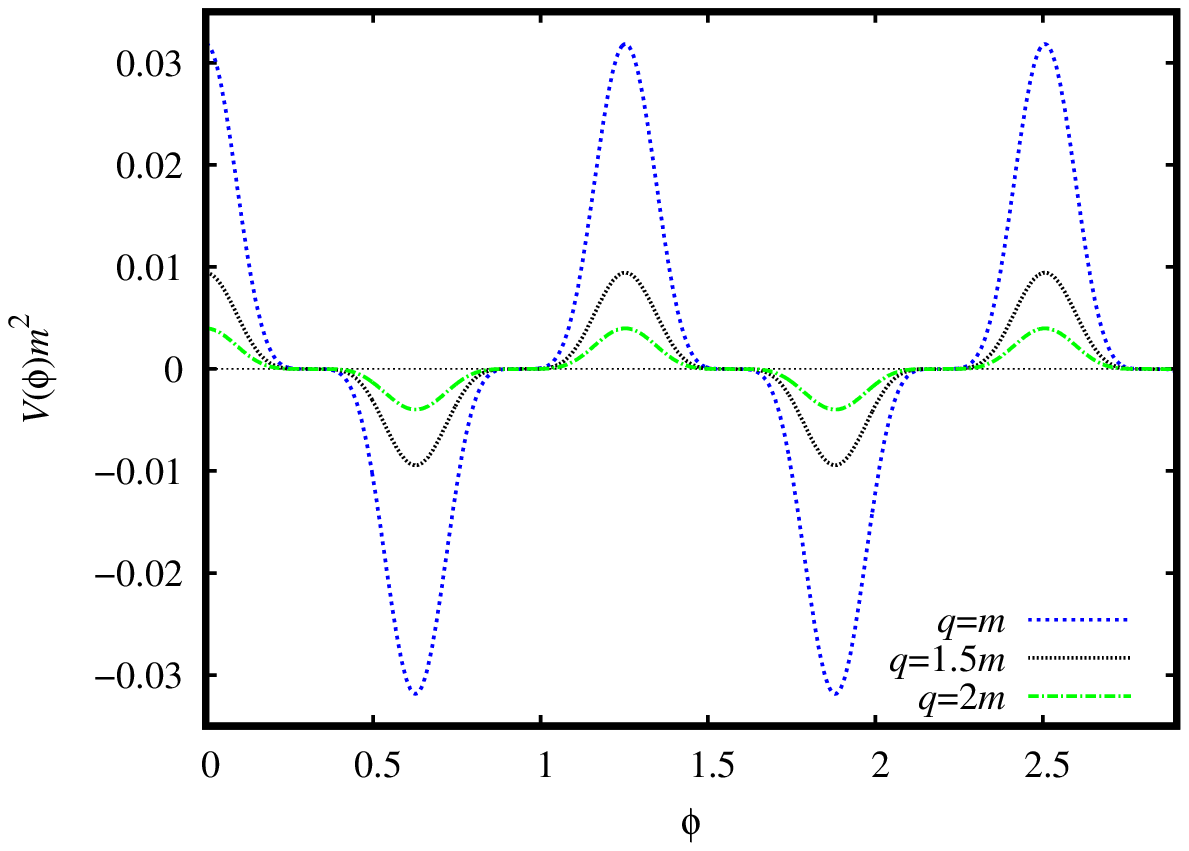}
    \includegraphics[scale=0.7]{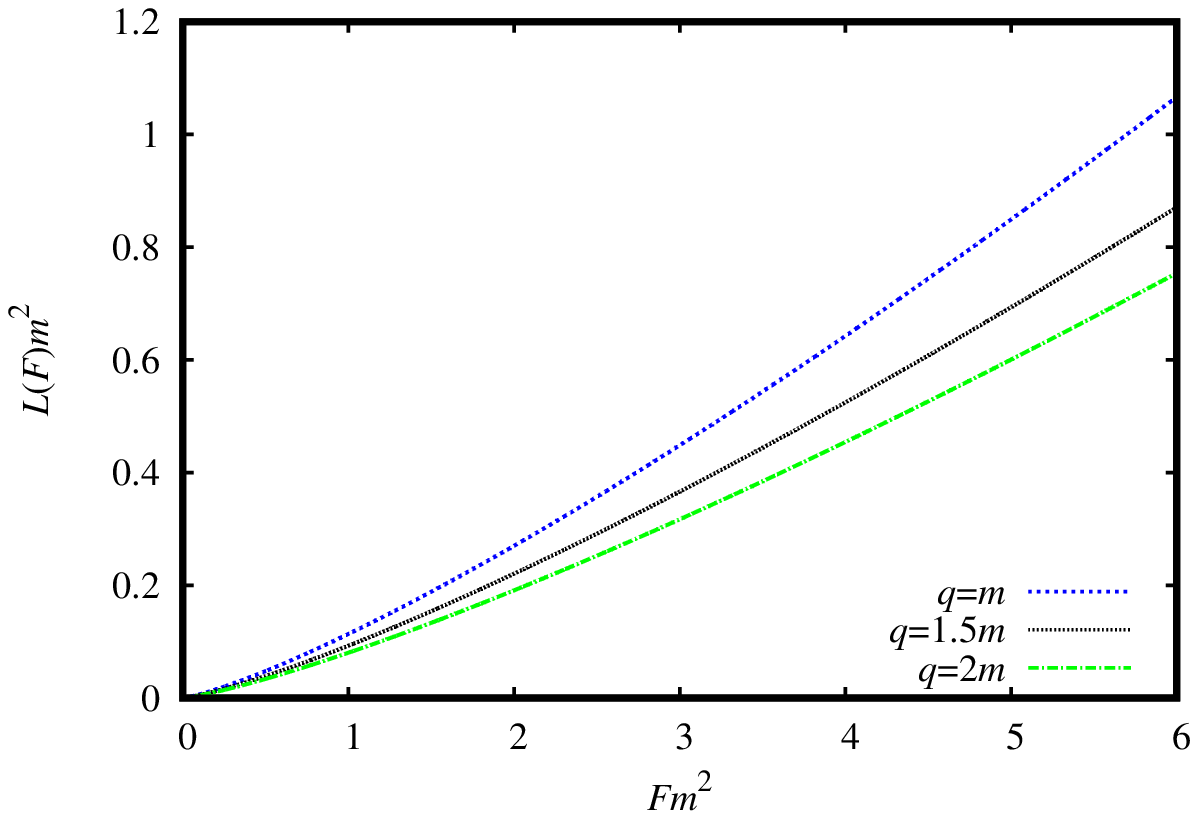}
    \caption{Behavior of the scalar field potential, as a function of $\phi$, and of the electromagnetic Lagrangian, as a function of $F$, that generates the Simpson--Visser solution to different values of charge.}
    \label{fig:VLSV}
\end{figure}

From \eqref{phi-SV} and \eqref{EMScalar}, we obtain the forms of $r(\phi)$ and $r(F)$. This allows us to get $V(\phi)$ and $L(F)$,
\begin{eqnarray}
    V(\phi)&=&\frac{4 m \cos ^5\left(\phi 
   \kappa \right)}{5 \kappa ^2 \left|q\right|^{3}},\label{Vphi-SV}\\
   L(F)&=&\frac{12 \sqrt[4]{2} m F^{5/4} }{5 \kappa ^2
   \sqrt{\left|q\right|}},\label{LLF-SV}
\end{eqnarray}
so that, we obtain the material content that describes the Simpson--Visser solution. In Fig. \ref{fig:VLSV} we see that the potential tends to a constant for a zero value of the scalar field and is periodic as $\phi$ increases. The intensity of the potential decreases for larger values of charge. We also see that the electromagnetic Lagrangian is not a multivalued function, as expected for a solution with a magnetic source, it tends to zero when $F=0$, and grows as $F$ increases.

\section{Bardeen-Type Black Bounce}\label{S:BT}
Now we consider the Bardeen-type black bounce solution described by the line element \eqref{lineel} with \cite{Lobo:2020ffi}
\begin{equation}
    f(r)=1-\frac{2mr^2}{\left(r^2+q^2\right)^{3/2}}, \quad \mbox{and} \quad \Sigma(r)=\sqrt{r^2+q^2}. \label{BDSOL}
\end{equation}

The scalar field is also given by equation \eqref{phi-SV}, since it depends on $\Sigma(r)$. The potential and the electromagnetic quantities are
\begin{eqnarray}
    V(r)&=&\frac{4 m \left(7 q^2 r^2-8 q^4\right)}{35 \kappa ^2
   \left(q^2+r^2\right)^{7/2}},\label{V-Bardeen}\\
   L(r)&=&\frac{2 m q^2 \left(16 q^2+91 r^2\right)}{35 \kappa ^2
   \left(q^2+r^2\right)^{7/2}},\label{L-Bardeen}\\
   L_F(r)&=&\frac{m \left(13 r^2-2 q^2\right)}{\kappa ^2 \left(q^2+r^2\right)^{3/2}}.\label{LF-Bardeen}
\end{eqnarray}
Equations \eqref{L-Bardeen} and \eqref{LF-Bardeen} satisfy the condition \eqref{EMbound}.

\begin{figure}
    \centering
    \includegraphics[scale=0.7]{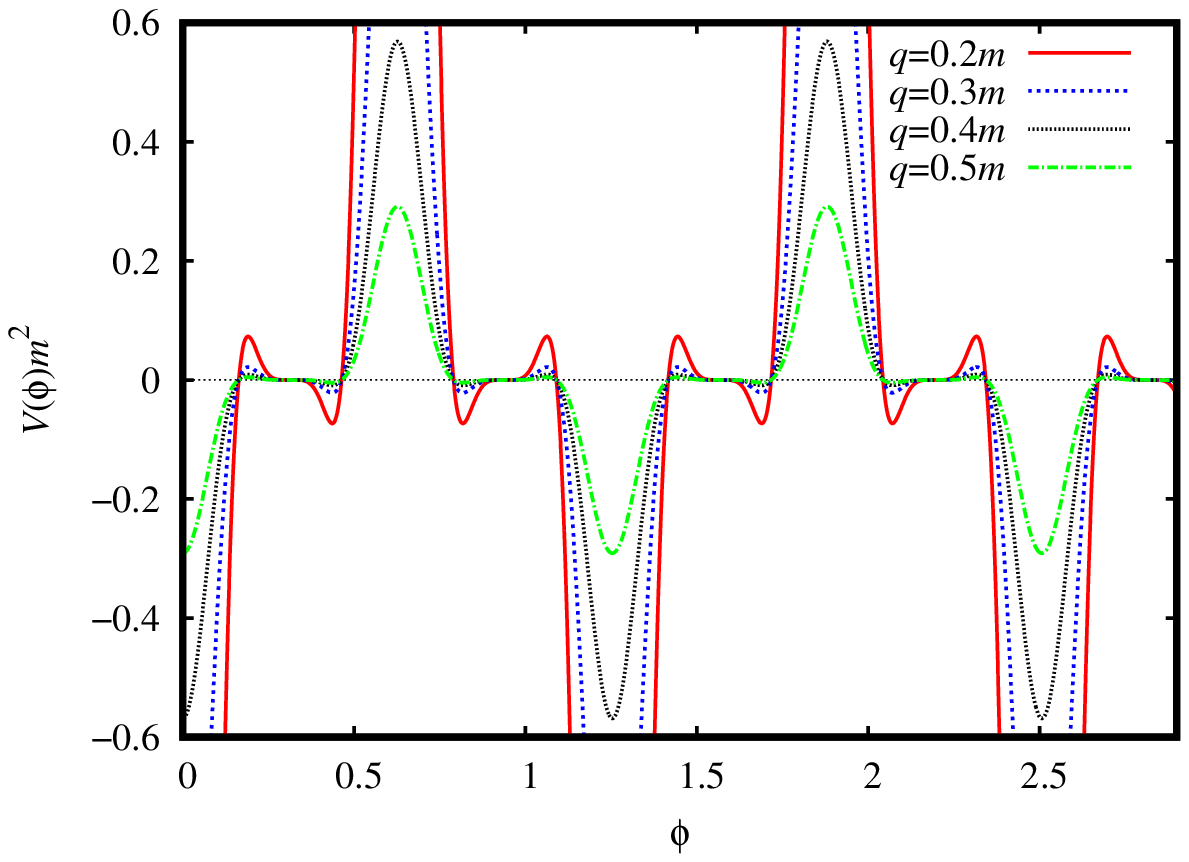}
    \includegraphics[scale=0.7]{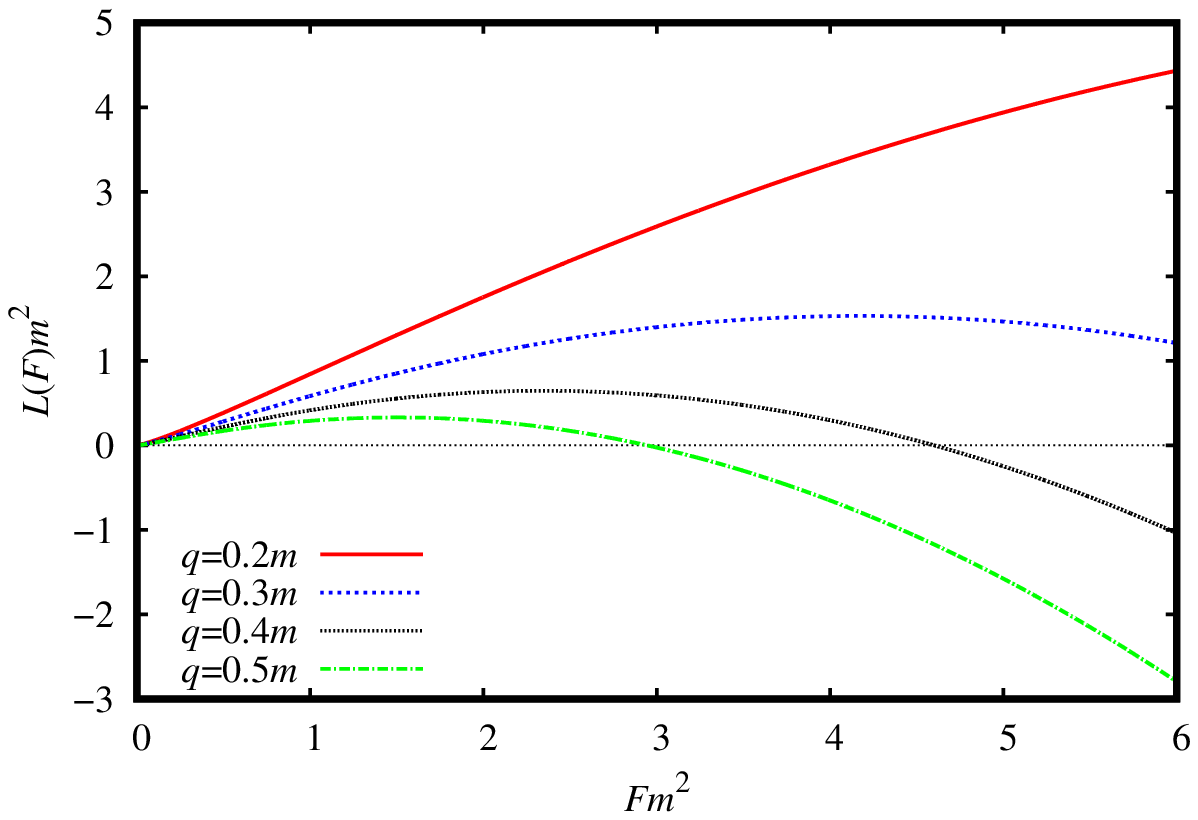}
    \caption{Behavior of the scalar field potential, as a function of $\phi$, and of the electromagnetic Lagrangian, as a function of $F$, that generates the Bardeen-type solution to different values of charge.}
    \label{fig:VLBD}
\end{figure}

Using \eqref{phi-SV} and \eqref{EMScalar}, we find $V(\phi)$ and $L(F)$, that are given by
\begin{eqnarray}
    V(\phi)&=&\frac{4 m \cos ^5\left(\phi \kappa\right) \left(7 \sin ^2\left(\phi \kappa\right)-8\cos^2\left(\phi \kappa\right)\right)}{35 \kappa ^2 \left|q^3 \right|},\label{Vphi-Bardeen}\\
    L(F)&=&\frac{4 \sqrt[4]{2} F^{5/4} m\left(91-75\sqrt{2F}
   q\right)}{35 \kappa ^2 \sqrt{\left|q\right|}}.\label{LLF-Bardeen}
\end{eqnarray}
With this, we realize that the Bardeen-type solution has more complications than the Simpson--Visser solution. The electromagnetic field does not behaves like Maxwell in the weak field limit. Through Fig. \ref{fig:VLBD}, we see graphically that the Bardeen-type solution has more complications in its matter content than the Simpson--Visser solution. The potential associated with the scalar field is periodic and in each cycle there are several maximums and minimums. The electromagnetic Lagrangian tends to zero to $F\rightarrow 0$. Unlike the Simpson--Visser case, the Lagrangian does not grow indefinitely with $F$. There is a maximum value that $L(F)$ can reach and after that it decreases, assuming negative values.

\section{Energy conditions}\label{S:EC}
As we said before, these black bounce solutions violate the null energy condition, such that the scalar field becomes real. Once the null energy condition is violated, all other conditions are violated. However, we can find out how the energy conditions behave separately for the scalar field and for the electromagnetic field. The energy conditions are given by
\begin{eqnarray}
&&NEC_{1,2}^{\phi,EM}=WEC_{1,2}^{\phi,EM}=SEC_{1,2}^{\phi,EM}
\Longleftrightarrow \rho^{\phi,EM}+p_{1,2}^{\phi,EM}\geq 0,\label{Econd1} \\
&&SEC_3 ^{\phi,EM}\Longleftrightarrow\rho^{\phi,EM}+p_1^{\phi,EM}+2p_2^{\phi,EM}\geq 0,\label{Econd2}\\
&&DEC_{1,2}^{\phi,EM} \Longrightarrow \rho^{\phi,EM}-p_{1,2}^{\phi,EM}\geq 0,\label{Econd3}\\
&&DEC_3^{\phi,EM}=WEC_3^{\phi,EM} \Longleftrightarrow\rho^{\phi,EM}\geq 0,\label{Econd4}
\end{eqnarray}
where the indexes $\phi$ and $EM$ determine whether the energy condition is associated with the scalar field or the electromagnetic field. The fluid quantities are identify through the stress-energy tensor as
\begin{eqnarray}
T^{\mu}{}_{\nu}={\rm diag}\left[\rho^{\phi,EM},-p_1^{\phi,EM},-p_2^{\phi,EM},-p_2^{\phi,EM}\right]\,,\label{EMT}
\end{eqnarray}
where $f(r)>0$. In regions where $f(r)<0$, the fluid quantities are identified as
\begin{eqnarray}
T^{\mu}{}_{\nu}={\rm diag}\left[-p_1^{\phi,EM},\rho^{\phi,EM},-p_2^{\phi,EM},-p_2^{\phi,EM}\right]\,.\label{EMTOut}
\end{eqnarray}
So that, in regions where $f(r)>0$, we have the following set of equations
\begin{eqnarray}
    \rho^{\phi}&=&-  f(r) \phi '(r)^2+V(r),\\
    p_1^{\phi}&=& - f(r) \phi '(r)^2-V(r),\\
    p_2^{\phi}&=&  f(r) \phi '(r)^2-V(r),\\
    \rho^{EM}&=&L(r),\\
    p_1^{EM}&=&-L(r),\\
    p_2^{EM}&=&-L(r)+\frac{q^2 L_F(r)}{\Sigma (r)^4}.
\end{eqnarray}
Finally, for regions where $f(r)<0$, we have
\begin{eqnarray}
    \rho^{\phi}&=&  f(r) \phi '(r)^2+V(r),\\
    p_1^{\phi}&=&  f(r) \phi '(r)^2-V(r),\\
    p_2^{\phi}&=&  f(r) \phi '(r)^2-V(r),\\
    \rho^{EM}&=&L(r),\\
    p_1^{EM}&=&-L(r),\\
    p_2^{EM}&=&-L(r)+\frac{q^2 L_F(r)}{\Sigma (r)^4}.
\end{eqnarray}
These equations allow us to obtain the energy conditions \eqref{Econd1}-\eqref{Econd4}, that, to $f(r)>0$, are given by
\begin{eqnarray}
&&NEC_{1}^{\phi}=WEC_{1}^{\phi}=SEC_{1}^{\phi}
\Longleftrightarrow -2 f(r) \phi '(r)^2\geq 0,\label{Econdphi1} \\
&&NEC_{2}^{\phi}=WEC_{2}^{\phi}=SEC_{2}^{\phi}
\Longleftrightarrow 0,\label{Econdphi2} \\
&&SEC_3 ^{\phi}\Longleftrightarrow -2 V(r)\geq 0,\label{Econdphi3}\\
&&DEC_{1}^{\phi} \Longrightarrow 2 V(r)\geq 0,\label{Econdphi4}\\
&&DEC_{2}^{\phi} \Longrightarrow 2 \left(V(r)-f(r) \phi '(r)^2\right)\geq 0,\label{Econdphi5}\\
&&DEC_3^{\phi}=WEC_3^{\phi} \Longleftrightarrow V(r)-f(r) \phi '(r)^2\geq 0,\label{Econdphi6}
\end{eqnarray}
\begin{eqnarray}
&&NEC_{1}^{EM}=WEC_{1}^{EM}=SEC_{1}^{EM}
\Longleftrightarrow 0,\label{EcondEM1} \\
&&NEC_{2}^{EM}=WEC_{2}^{EM}=SEC_{2}^{EM}
\Longleftrightarrow \frac{q^2 {L_F}(r)}{\Sigma (r)^4}\geq 0,\label{EcondEM2} \\
&&SEC_3 ^{EM}\Longleftrightarrow \frac{2 q^2 {L_F}(r)}{\Sigma (r)^4}-2 L(r)\geq 0,\label{EcondEM3}\\
&&DEC_{1}^{EM} \Longrightarrow 2 L(r)\geq 0,\label{EcondEM4}\\
&&DEC_{2}^{EM} \Longrightarrow 2 L(r)-\frac{q^2 {L_F}(r)}{\Sigma (r)^4}\geq 0,\label{EcondEM5}\\
&&DEC_3^{EM}=WEC_3^{EM} \Longleftrightarrow L(r)\geq 0.\label{EcondEM6}
\end{eqnarray}
From  \eqref{Econdphi2}, we see that $NEC_{2}^{\phi}$ is identically satisfied to the scalar case and from \eqref{EcondEM1} we see that $NEC_{1}^{EM}$ is also identically satisfied. For regions where $f(r)<0$, the structure of the energy conditions for the electromagnetic sector does not change, remaining equal to equations \eqref{EcondEM1}-\eqref{EcondEM6}. To the scalar sector, we find
\begin{eqnarray}
&&NEC_{1}^{\phi}=WEC_{1}^{\phi}=SEC_{1}^{\phi}
\Longleftrightarrow 2 f(r) \phi '(r)^2\geq 0,\label{Econdphi11} \\
&&NEC_{2}^{\phi}=WEC_{2}^{\phi}=SEC_{2}^{\phi}
\Longleftrightarrow 2V(r)\geq 0,\label{Econdphi22} \\
&&SEC_3 ^{\phi}\Longleftrightarrow 2 V(r)\geq 0,\label{Econdphi33}\\
&&DEC_{1}^{\phi} \Longrightarrow 2 V(r)\geq 0,\label{Econdphi44}\\
&&DEC_{2}^{\phi} \Longrightarrow2 f(r) \phi '(r)^2\geq 0,\label{Econdphi55}\\
&&DEC_3^{\phi}=WEC_3^{\phi} \Longleftrightarrow V(r)+f(r) \phi '(r)^2\geq 0.\label{Econdphi66}
\end{eqnarray}
Equations \eqref{Econdphi1}-\eqref{Econdphi6} are essentially different, for the most part, from equations \eqref{Econdphi11}-\eqref{Econdphi66}. Now we have the requirements to evaluate the energy conditions for the models studied in this work.

\subsection{Simpson--Visser solution}
To the Simpson--Visser model, we replace equations \eqref{phi-SV}-\eqref{LF-SV} in the energy conditions, that results, to $f(r)>0$ in
\begin{eqnarray}
&&NEC_{1}^{\phi}
\Longleftrightarrow -\frac{2 q^2 \left(\sqrt{q^2+r^2}-2 m\right)}{\kappa ^2 \left(q^2+r^2\right)^{5/2}}\geq 0, \qquad SEC_3 ^{\phi}\Longleftrightarrow -\frac{8 m q^2}{5 \kappa ^2 \left(q^2+r^2\right)^{5/2}}\geq 0,\label{EcondphiSV1}\\
&&DEC_{1}^{\phi} \Longrightarrow \frac{8 m q^2}{5 \kappa ^2 \left(q^2+r^2\right)^{5/2}}\geq 0,\qquad
DEC_{2}^{\phi} \Longrightarrow -\frac{2 q^2 \left(5 \sqrt{q^2+r^2}-14 m\right)}{5 \kappa ^2 \left(q^2+r^2\right)^{5/2}}\geq 0,\label{EcondphiSV2}\\
&&WEC_3^{\phi} \Longleftrightarrow -\frac{q^2 \left(5 \sqrt{q^2+r^2}-14 m\right)}{5 \kappa ^2 \left(q^2+r^2\right)^{5/2}}\geq 0,\label{EcondphiSV6}
\end{eqnarray}

\begin{eqnarray}
&&NEC_{2}^{EM}
\Longleftrightarrow \frac{3 m q^2}{\kappa ^2 \left(q^2+r^2\right)^{5/2}}\geq 0,\qquad
SEC_3 ^{EM}\Longleftrightarrow \frac{18 m q^2}{5 \kappa ^2 \left(q^2+r^2\right)^{5/2}}\geq 0,\label{EcondEMSV1}\\
&&DEC_{1}^{EM} \Longrightarrow \frac{12 m q^2}{5 \kappa ^2 \left(q^2+r^2\right)^{5/2}}\geq 0,\qquad
DEC_{2}^{EM} \Longrightarrow -\frac{3 m q^2}{5 \kappa ^2 \left(q^2+r^2\right)^{5/2}}\geq 0,\label{EcondEMSV2}\\
   &&WEC_3^{EM} \Longleftrightarrow \frac{6 m q^2}{5 \kappa ^2 \left(q^2+r^2\right)^{5/2}}\geq 0.\label{EcondEMSV3}
\end{eqnarray}
If there is an event horizon, outside the horizon, the scalar field will not violate the inequalities $NEC_2^{\phi}$ and $DEC_1^{\phi}$. The scalar field violates all the energy conditions in this region. The electromagnetic field violates only the dominant energy condition once only the inequality $DEC_2^{EM}$ is not satisfied.

Inside the possible event horizon, we find
\begin{eqnarray}
&&NEC_{1}^{\phi}
\Longleftrightarrow \frac{2 q^2 \left(\sqrt{q^2+r^2}-2 m\right)}{\kappa ^2 \left(q^2+r^2\right)^{5/2}}\geq 0,\qquad NEC_{2}^{\phi}
\Longleftrightarrow \frac{8 m q^2}{5 \kappa ^2 \left(q^2+r^2\right)^{5/2}}\geq 0,\label{EcondphiSV11} \\
&&SEC_3 ^{\phi}\Longleftrightarrow \frac{8 m q^2}{5 \kappa ^2 \left(q^2+r^2\right)^{5/2}}\geq 0,\qquad WEC_3^{\phi} \Longleftrightarrow \frac{q^2 \left(5 \sqrt{q^2+r^2}-6 m\right)}{5 \kappa ^2 \left(q^2+r^2\right)^{5/2}}\geq 0,\label{EcondphiSV22}\\
&&DEC_{1}^{\phi} \Longrightarrow \frac{8 m q^2}{5 \kappa ^2 \left(q^2+r^2\right)^{5/2}}\geq 0,\qquad
DEC_{2}^{\phi} \Longrightarrow\frac{2 q^2 \left(\sqrt{q^2+r^2}-2 m\right)}{\kappa ^2 \left(q^2+r^2\right)^{5/2}}\geq 0.\label{EcondphiSV33}
\end{eqnarray}
The scalar field violates all energy condition once the inequalities $NEC_{1}^{\phi}$, $WEC_{3}^{\phi}$, and $DEC_{2}^{\phi}$ are not satisfied.

\subsection{Bardeen-type solution}
To the Bardeen-type model, we replace equations \eqref{phi-SV}, and \eqref{V-Bardeen}-\eqref{LF-Bardeen} in the energy conditions, that results, to $f(r)>0$ in
\begin{eqnarray}
&&NEC_{1}^{\phi}
\Longleftrightarrow -\frac{2 q^2\left(1-\frac{2 m r^2}{\left(q^2+r^2\right)^{3/2}}\right)}{\kappa ^2 q^2
   \left(r^2+q^2\right)^2}\geq 0, \qquad SEC_3 ^{\phi}\Longleftrightarrow -\frac{8 m q^2 \left(8 q^2-7 r^2\right)}{35 \kappa ^2 \left(q^2+r^2\right)^{7/2}}\geq 0,\label{EcondphiBD1}\\
&&DEC_{1}^{\phi} \Longrightarrow \frac{8 m \left(7 q^2 r^2-8 q^4\right)}{35 \kappa ^2 \left(q^2+r^2\right)^{7/2}}\geq 0,\qquad
DEC_{2}^{\phi} \Longrightarrow -\frac{2 q^2 \left(5 \sqrt{q^2+r^2}-14 m\right)}{5 \kappa ^2 \left(q^2+r^2\right)^{5/2}}\geq 0,\label{EcondphiBD2}\\
&&WEC_3^{\phi} \Longleftrightarrow -\frac{q^2 \left(32 m q^2-98 m r^2+35 \left(q^2+r^2\right)^{3/2}\right)}{35 \kappa ^2
   \left(q^2+r^2\right)^{7/2}}\geq 0,\label{EcondphiBD3}
\end{eqnarray}

\begin{eqnarray}
&&NEC_{2}^{EM}
\Longleftrightarrow \frac{m q^2 \left(13 r^2-2 q^2\right)}{\kappa ^2 \left(q^2+r^2\right)^{7/2}}\geq 0,\qquad
SEC_3 ^{EM}\Longleftrightarrow -\frac{6 m q^2 \left(34 q^2-91 r^2\right)}{35 \kappa ^2 \left(q^2+r^2\right)^{7/2}}\geq 0,\label{EcondEMBD1}\\
&&DEC_{1}^{EM} \Longrightarrow \frac{4 m q^2 \left(16 q^2+91 r^2\right)}{35 \kappa ^2 \left(q^2+r^2\right)^{7/2}}\geq 0,\qquad
DEC_{2}^{EM} \Longrightarrow \frac{m q^2 \left(134 q^2-91 r^2\right)}{35 \kappa ^2 \left(q^2+r^2\right)^{7/2}}\geq 0,\label{EcondEMBD2}\\
   &&WEC_3^{EM} \Longleftrightarrow \frac{2 m q^2 \left(16 q^2+91 r^2\right)}{35 \kappa ^2 \left(q^2+r^2\right)^{7/2}}\geq 0.\label{EcondEMBD3}
\end{eqnarray}
To the scalar field, where $f(r)>0$, the inequalities $NEC_{1}^{\phi}$, $DEC_{2}^{\phi}$, and $WEC_{3}^{\phi}$ are not satisfied, so that, all energy conditions are violated. The electromagnetic field violates the dominant energy condition once the inequality $DEC_2^{EM}$ is not satisfied to $r>>1$. The conditions $SEC_3^{EM}$ and $NEC_2^{EM}$ will not necessarily be violated outside of a possible event horizon. This will depend on the charge value of the solution. However, even if these conditions are not violated outside the horizon, they are violated within the horizon.

In the region where $f(r)<0$, we find
\begin{eqnarray}
&&NEC_{1}^{\phi}
\Longleftrightarrow \frac{2 q^2\left(1-\frac{2 m r^2}{\left(q^2+r^2\right)^{3/2}}\right)}{\kappa ^2 q^2
   \left(r^2+q^2\right)^2}\geq 0,\qquad NEC_{2}^{\phi}
\Longleftrightarrow \frac{8 m \left(7 q^2 r^2-8 q^4\right)}{35 \kappa ^2 \left(q^2+r^2\right)^{7/2}}\geq 0,\label{EcondphiBD11} \\
&&SEC_3 ^{\phi}\Longleftrightarrow \frac{8 m q^2 \left(7 r^2-8 q^2\right)}{35 \kappa ^2 \left(q^2+r^2\right)^{7/2}}\geq 0,\qquad WEC_3^{\phi} \Longleftrightarrow \frac{q^2 \left(-32 m q^2-42 m r^2+35 \left(q^2+r^2\right)^{3/2}\right)}{35 \kappa ^2
   \left(q^2+r^2\right)^{7/2}}\geq 0,\label{EcondphiBD22}\\
&&DEC_{1}^{\phi} \Longrightarrow \frac{8 m q^2 \left(7 r^2-8 q^2\right)}{35 \kappa ^2 \left(q^2+r^2\right)^{7/2}}\geq 0,\qquad
DEC_{2}^{\phi} \Longrightarrow \frac{2 q^2\left(1-\frac{2 m r^2}{\left(q^2+r^2\right)^{3/2}}\right)}{\kappa ^2 q^2
   \left(r^2+q^2\right)^2}\geq 0.\label{EcondphiBD33}
\end{eqnarray}
The scalar field violates all energy condition once the null energy condition is not satisfied.

\section{Conclusion}\label{S:conclusion}
In this work, we obtain the material content of black bounce solutions in general relativity. For this, we consider the coupling of gravitational theory with a scalar field and nonlinear electrodynamics.

We show that, as the stress-energy tensor of black bounces does not satisfy certain symmetries, only nonlinear electrodynamics is not enough to generate these solutions. For this reason, we consider the theory described by the action \eqref{action} and obtained the field equations for this theory. The parameter $\epsilon$ determines whether we have the presence of an usual scalar field or a phantom scalar field. Through the field equations, we built the necessary formalism to obtain the material content to a general solution. With these results, we propose a theorem that relates the need for the scalar field to be phantom with the fact that the null energy condition is violated. It means that, if the null energy condition were satisfied, the scalar field that generates the black bounce could be nonphantom. 

We applied the method to the Simpson--Visser solution and the Bardeen-type black bounce solution and obtained the shape of the Lagrangian $L(F)$, equations \eqref{LLF-SV} and \eqref{LLF-Bardeen}, and the potential $V(\phi)$, equations \eqref{Vphi-SV} and \eqref{Vphi-Bardeen}, that generate these solutions. The functions \eqref{L-SV}, \eqref{LF-SV}, \eqref{L-Bardeen}, and \eqref{LF-Bardeen} are obtained independently through the field equations. However, they must satisfy the condition \eqref{EMbound}, what really happens.

Once we have separately the stress-energy tensor of each field that composes the source, we analyze the energy conditions associated with each field. In the case of the Simpson--Visser Solution, the scalar field violated all energy conditions, while the electromagnetic field violated only the dominant energy condition. In the case of the Bardeen-type solution, both the scalar field and the electromagnetic field violate all energy conditions. However, for the electromagnetic field, it is possible to at least guarantee the positivity of the energy density, $\rho^{EM}=WEC_3^{EM}$.

There are several black bounce solutions. However, it will not always be possible to apply the method presented in this work to obtain the material content in an analytical way. In some cases, it is not possible to solve equation \eqref{Vgeneral} analytically.

Now that we have the material content of these solutions, we can analyze other properties associated with the material content, such as thermodynamics and perturbations.


\section*{Acknowledgments}
M.E.R.  thanks Conselho Nacional de Desenvolvimento Cient\'ifico e Tecnol\'ogico - CNPq, Brazil  for partial financial support. 


\end{document}